\documentclass[usenatbib,useAMS]{mnras}
\usepackage{tablefootnote}
\usepackage{multirow}
\usepackage{xcolor}
\usepackage{hyperref}
\usepackage{graphicx}	
\usepackage{amsmath}	
\usepackage{amssymb}	
\usepackage{multicol}        
\usepackage{bm}		
\usepackage{caption}

\title[4.7 and 8.2 GHz PSO J0309+2717 observations]{Flux-density measurements of the high-redshift blazar PSO J047.4478+27.2992 at 4.7 and 8.2 GHz with RATAN-600} 

\author[T. Mufakharov et al.]{
T.~Mufakharov,$^{1,2}$\thanks{E-mail: timur.mufakharov@gmail.com} 
A. Mikhailov,$^{3}$
Yu. Sotnikova,$^{3}$
M. Mingaliev,$^{1,3}$
V. Stolyarov,$^{1,3,4}$
A. Erkenov,$^{3}$
\newauthor
N. Nizhelskij$^{3}$
and  P. Tsybulev$^{3}$
\\
$^{1}$Kazan Federal University, 18 Kremlyovskaya St, Kazan 420008, Russia\\
$^{2}$Shanghai Astronomical Observatory, Chinese Academy of Sciences, Shanghai 200030, China\\
$^{3}$Special Astrophysical Observatory of RAS, Nizhny Arkhyz 369167, Russia\\
$^{4}$Astrophysics Group, Cavendish Laboratory, University of Cambridge,
      J J Thomson Avenue, Cambridge CB3 0HE, UK 
}

\date{Accepted 2020 November 21. Received 2020 November 20; in original form 2020 August 18}
\pubyear{2020}
\begin{document}
\label{firstpage}
\pagerange{\pageref{firstpage}--\pageref{lastpage}}
\maketitle

\begin{abstract}
We report the first detection at 4.7 and 8.2 GHz with the RATAN-600 radio telescope of the source PSO J047.4478+27.2992, which has been announced as a blazar at $z=6.1$. The average flux densities are $12\pm3$ and $8\pm4$ mJy at 4.7 and 8.2 GHz respectively, and an upper limit is estimated as 3 mJy at 11.2 GHz. The light curve at 4.7 GHz, obtained on a time-scale of four months, exhibits moderate variability of the source (fractional variability $F_{var}=0.28\pm0.02$).
The new RATAN-600 measurements extend previous literature data to higher frequencies, indicating a single power-law radio spectrum
 with $\alpha_{0.147-8.2}$=$-0.51\pm0.1$. The monochromatic radio luminosity at 4.7 GHz $P_{\nu} \sim 2 \times 10^{27}$ W Hz$^{-1}$ is consistent with values for high-redshift quasars at $z\geq 3$.  
\end{abstract}



\section{Introduction}
The newly discovered high-redshift blazar PSO J047.4478+27.2992 
 at $z=6.1$, also known as NVSS J030947+271757 
 and as PSO J030947.49+271757.31 in \cite{2020A&A...635L...7B}, 
 is the only known blazar at $z>6$.
In this paper, we will refer to the object as  PSO J0309+2717. 
Its redshift was obtained from spectroscopic observations 
 with the Large Binocular Telescope in 2019. 
 The radio, optical, and X-ray photometric and spectroscopic properties discussed in 
 \cite{2020A&A...635L...7B} are the basis of the hypothesis that its radio and X-ray emissions are jet dominated.
 The object was observed in X-rays in the 0.5-10 keV energy band with \textit{Swift/XRT}, 
and it appears to be 
 the second most luminous active galactic nucleus (AGN) in X-rays at $z>6$ \citep{2020A&A...635L...7B, 2020MNRAS.497.1842M}.
 It has three measurements 
 in the radio band: $23.89 \pm 0.87$ mJy at 1.4 GHz (NVSS, \citealt{1998AJ....115.1693C}),
  $64.2 \pm 6.2$ mJy at 147 MHz (TGSS, \citealt{2017A&A...598A..78I}),
 and $11.76 \pm 0.23$ mJy at 3 GHz (VLASS\footnote{\url{https://cirada.ca/catalogues}}, \citealt{2020PASP..132c5001L,2020RNAAS...4..175G}).

Blazars are radio-loud AGNs with their relativistic jets seen at a small angle to the observer's line of sight \citep{1995PASP..107..803U};
this is why they can be visible at high redshifts and provide information about the early Universe. 
 However, the fraction of radio-loud quasars decreases with increasing 
 redshift starting from $z = 3$ \citep{2014AJ....147....6M}.
 This could be explained by the change in the relative accretion modes
 and spin magnitude of supermassive black holes (SMBHs) in AGNs \citep{2007ApJ...656..680J, 2007AJ....134..617W, 2013ApJ...762...68D},
 because SMBHs need a significant amount of time to form,
 and this happens in the first $10^9$ yr after the Big Bang \citep{2015eaci.book.....S}.
At high redshifts ($z > 3$) fewer quasars are expected to be found because they are either not fully formed or have not reached a peak of activity yet; 
therefore it is hard to detect them \citep{1995AJ....110...68S, 2014Madau}.
The estimated radio-loud fraction of quasars is between 6 and 19 per cent at $z\sim6$ \citep{2015ApJ...804..118B}. This fraction argues against the strong evolution of radio-loud quasars with redshift. As an example, there are 106 radio-loud quasars at redshifts $z\geq 3$ in the sky area $00h < RA <24h$, $-34\degr <  Dec. < 49\degr$ with flux densities $S \geq 100$ mJy at 1.4 GHz, and only 11 at  $z\geq 4$ (NASA/IPAC Extragalactic Database). 
There are very few measurements for them in other frequency bands.

Here we present the results of simultaneous RATAN-600 observations of PSO J0309+2717 at three frequencies  (4.7,  8.2,  and 11.2 GHz) from 2020 May--September.
Our goal was to measure the radio continuum spectrum of the object at frequencies higher than 1.4 GHz and evaluate its variability properties in the radio domain.
 
\section{RATAN-600 observations and results}
 
The RATAN-600 radio telescope was used in the transit mode at six frequencies from 1.2 to 22.3 GHz \citep{1993IAPM...35....7P}. 
The parameters of the antenna and receivers are listed in Table~\ref{tab:radiometers}
 (a detailed description is given in e.g. \citealt{2017AN....338..700M}).

 Data reduction and calibration were done using the RATAN standard methods \citep{1985BSAO...19...59A,2016AstBu..71..496U}. 
 The observations were spread over 4, 2, and 45-d periods in May, June, and July--September respectively
 (epoch 1, epoch 2, and epoch 3), for which
  one-dimensional scans were averaged to increase the detection level. 
 As a result, we obtained flux densities for three observational epochs (Table~\ref{table:fluxes}). 

The average signal at 4.7 GHz was measured with a signal-to-noise (S/N) level $>4\sigma$ for every epoch,  
 and the flux densities are
 $14\pm3$ mJy for epoch 1 and $10\pm2$ mJy for epoch 3.
 At the 8.2 GHz frequency a flux density of $8\pm3$ mJy
 was measured only for epoch 3 at a level $>2\sigma$ (Table~\ref{table:fluxes}). 
 We estimated an upper limit of 3 mJy for the 11.2 GHz flux density.
 
The total relative rms error of each flux-density measurement $\sigma_{S}/S_{\nu}$ 
is estimated using the relation \citep{2016AstBu..71..496U}.
\begin{equation}
(\frac{\sigma_{S}}{S_{\nu}})^2 = (\frac{\sigma_{c}}{g_{\nu}(e)})^2 + (\frac{\sigma_{m}}{T_{ant,\nu}})^2
\end{equation}
where $\sigma_{S}$ is the total standard flux-density error;
$S_{\nu}$ the flux density at a frequency ${\nu}$;
$\sigma_{c}$ the standard calibration curve error, which is about 1-2 per cent and 2-5 per cent at 4.7 and 8.2 GHz respectively;
$g_{\nu}(e)$ the elevation calibration function;
$\sigma_{m}$ the standard error of the antenna temperature $T_{ant}$ measurement;
and $T_{ant,\nu}$ the antenna temperature.

The uncertainty of the antenna temperature measurement depends on the receiver's noise, the atmospheric fluctuations, 
 and the accuracy of the antenna surface setting for an actual source observation. The systematic uncertainty of the absolute flux-density scale (3--7 per cent at  frequencies of 4.7 and 8.2 GHz) is not included in the total flux error. The range of rms errors of single flux-density measurements is 20--60 per cent for PSO J0309+2717 at 4.7 GHz. The total flux-density errors at 4.7 and 8.2 GHz for averaged scans (see Table \ref{table:fluxes}) are 22 per cent and 37 per cent respectively.
  
\begin{figure}
   \centering
   \includegraphics[width=8.5cm]{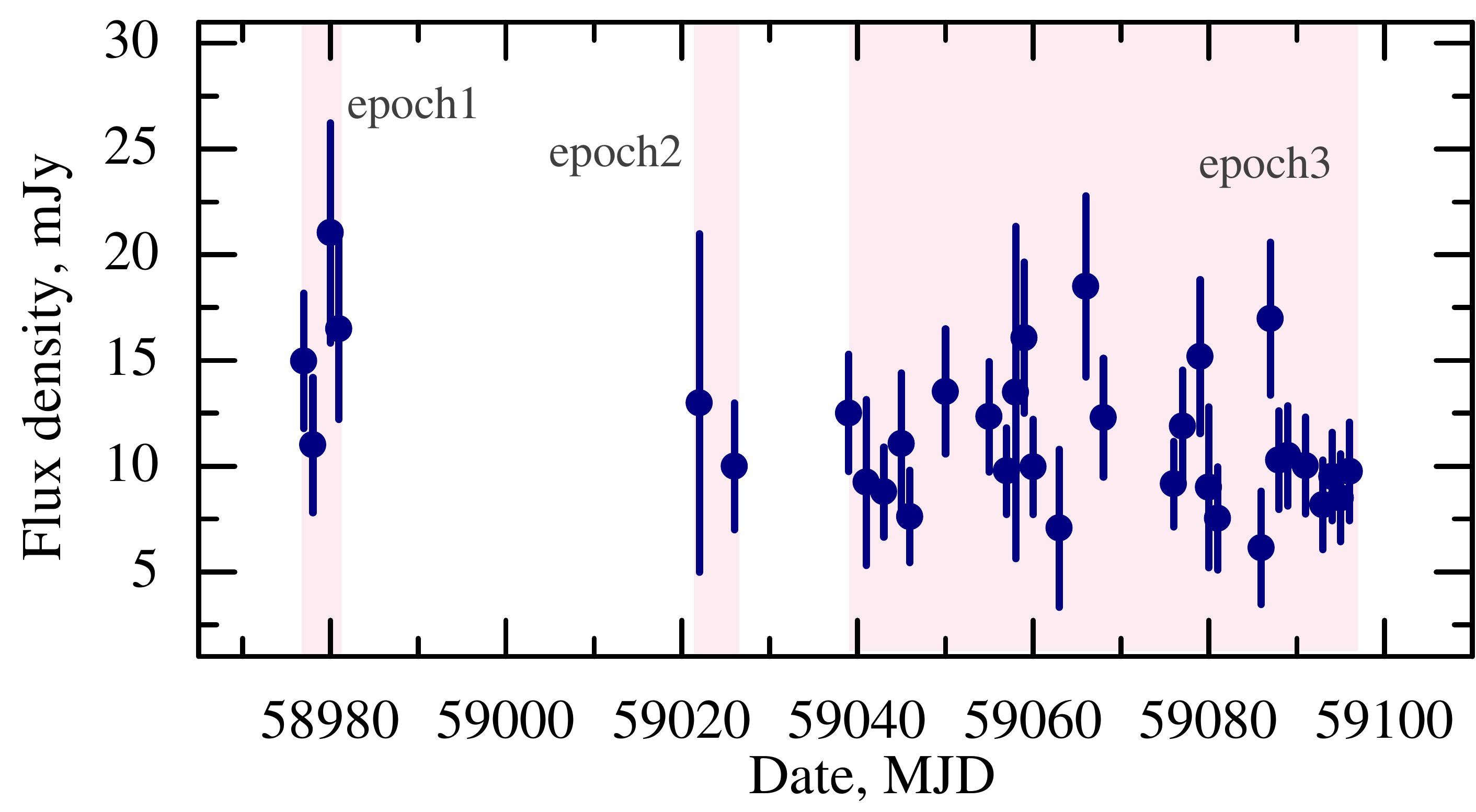}
      \caption{Light curve of PSO J0309+2717 based on every 4.7 GHz flux-density measurement during three epochs of observations (2020 May--September).}
         \label{Fig:figure1}
   \end{figure}

 The light curve of PSO J0309+2717 at 4.7 GHz is constructed using measurements with S/N $ >1.5\sigma$ (Fig~\ref{Fig:figure1}). 
 The variability estimates are of the same order as the measurement uncertainties: the variability index 
 $V = 0.23$ and the fractional variability $F_{var} = 0.28 \pm 0.02$. We used the following formulas to calculate them:
 
 \begin{equation}
V_{S}=\frac{(S_{max}-\sigma_{S_{max}})-(S_{min}+\sigma_{S_{min}})}
{(S_{max}-\sigma_{S_{max}})+(S_{min}+\sigma_{S_{min}})},
\end{equation}
where $S_{max}$ and $S_{min}$ are the maximum and minimum flux densities, and 
 $\sigma_{S_{max}}$ and $\sigma_{S_{min}}$ are their errors \citep{1992ApJ...399...16A}, and  

 \begin{equation}
F_{var} = \sqrt{\frac{V^2 - \bar{\sigma_{err}^2}}{\bar{S}^2}}, 
\end{equation}
where $V^2$ is the variance, $\bar{S}$ is the mean value of the flux density, and $\sigma_{err}$ the root mean square error \citep{2003MNRAS.345.1271V}.

The uncertainty of the fractional variability is determined as:
 \begin{equation}
\Delta F_{var} = \sqrt{ (\sqrt{\frac{1}{2N}} \frac{\bar{\sigma_{err}^2}}{F_{var}*\bar{S}^2})^2 + (\sqrt{\frac{\bar{\sigma_{err}^2}}{N}} \frac{1}{\bar{S}^2})^2}. 
\end{equation}

\begin{figure}
   \centering
   \includegraphics[width=8.5cm]{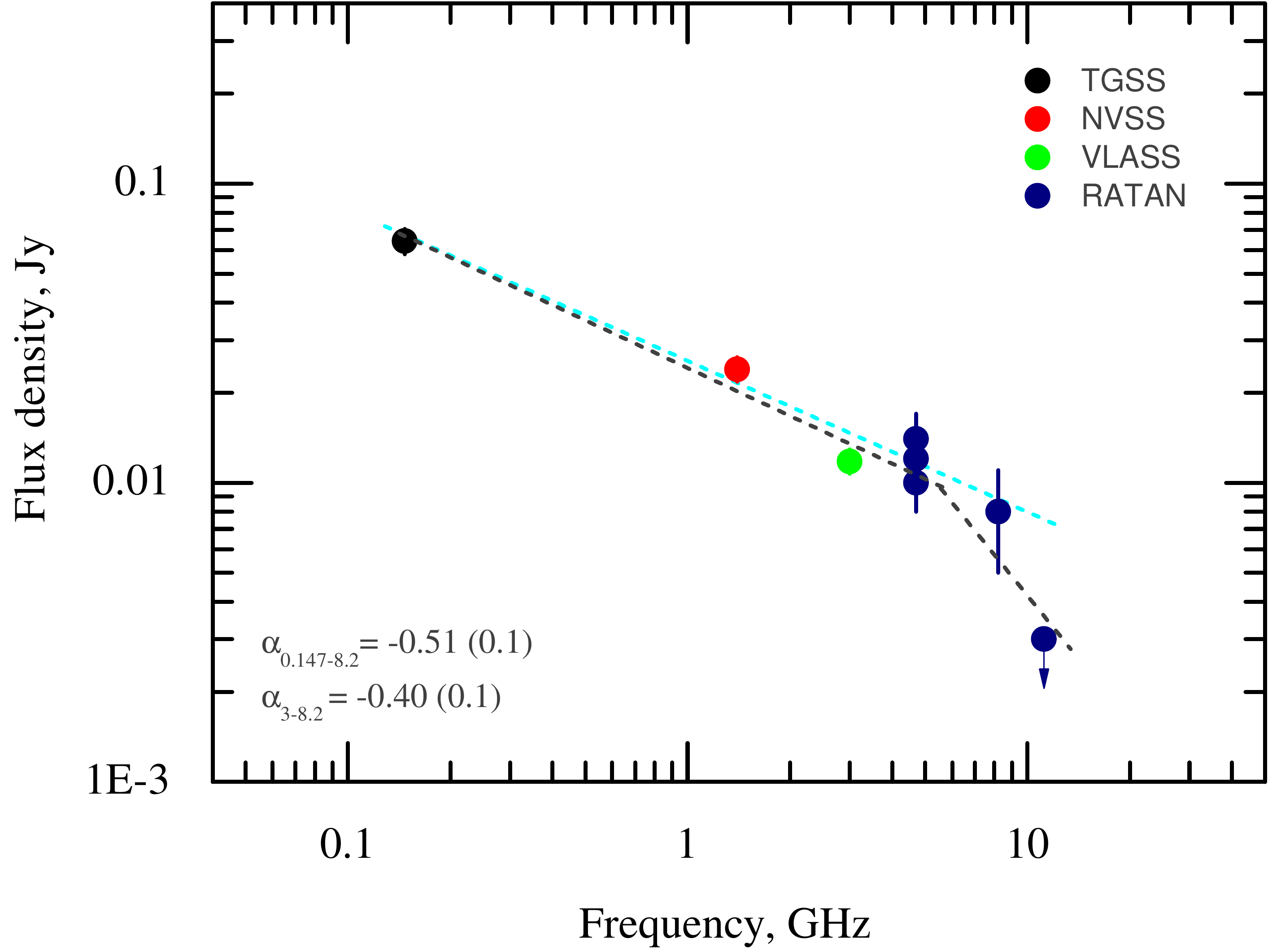}
      \caption{Radio continuum spectrum of PSO J0309+2717 constructed using both literature data at 0.147, 1.4, and 3 GHz
       and RATAN-600 simultaneous measurements at 4.7/8.2 GHz. } 
         \label{Fig:figure2}
   \end{figure}

The radio spectrum of the source is shown in Fig.~\ref{Fig:figure2} and 
summarizes the available radio measurements of NVSS, TGSS, VLASS, and RATAN.
In the case of the VLASS data, we adopted the flux-density value as stated in the current version of the catalogue,
 although the authors note that it is likely underestimated by $\sim$10 per cent. In the latter case, the 3 GHz data point
 in the radio spectrum would be slightly higher.
Assuming a single power-law distribution for continuum emission, 
the average radio spectral index $S\sim\nu^{\alpha}$
was calculated as $\alpha_{0.147-8.2} = -0.51\pm0.1$.
If we assume low or moderate variability at 3, 4.7, and 8.2 GHz on a time-scale of 1.5 years (2019 March--2020 September), 
the spectral index calculated based on the VLASS and RATAN data $\alpha_{3-8.2} = -0.40 \pm 0.1$ could be considered as quasi-simultaneous.
Considering the upper limit at 11.2 GHz, the spectrum of PSO J0309+2717 steepens at high frequencies ($ > 8$ GHz), and if we fit the spectrum with two linear components, we obtain flat and ultra-steep spectrum parts with $\alpha_{low}=-0.53 \pm 0.02$ and $\alpha_{high} \leq -1.4 \pm 0.05$. This indicates that the object can be a compact steep-spectrum (CSS) or a megahertz-peaked-spectrum (MPS) source with a maximum flux density at frequencies less than 0.147 GHz (1 GHz or less in the rest frame).

We estimated the monochromatic radio luminosity at 4.7 GHz $P_{\nu} \sim 2\times 10^{27}$ W Hz$^{-1}$.
We used the $\Lambda$CDM cosmology with $H_0 = 67.74$ km s$^{-1}$ Mpc$^{-1}$, $\Omega_m$=0.3089, 
and $\Omega_\Lambda$=0.6911 \citep{2016A&A...594A..13P}, and followed the standard formula:
 \begin{equation}
P_{\nu} = 4 \pi D_{L}^2 S_{\nu} (1+z)^{-\alpha -1} 
\end{equation}
where $\nu$ is the frequency, $S_{\nu}$ is the measured flux density, 
$z$ is the redshift, $\alpha$ is the spectral index, and $D_{L}$ is the luminosity distance. 
 
\begin{table}
\caption{RATAN-600 continuum radiometer parameters: the central frequency $f_0$, the bandwidth $\Delta f_0$,
the detection limit for point sources per transit $\Delta F$. FWHM$_{RA \times Dec.}$ is the angular resolution for RA and Dec., calculated for $\delta = 27\degr$.} 
\label{tab:radiometers}
\centering
\begin{tabular}{cccc}
\hline
 $f_{0}$ & $\Delta f_{0}$ & $\Delta F$ &  FWHM$_{RA \times Dec.}$\\
  GHz    &   GHz           &  mJy/beam   &    \\
\hline
 $22.3$ & $2.5$  &  $50$ & $0.14\arcmin \times 0.8\arcmin$  \\ 
 $11.2$ & $1.4$  &  $15$ & $0.28\arcmin \times 1.64\arcmin$ \\ 
 $8.2$  & $1.0$  &  $10$ & $0.39\arcmin \times 2.43\arcmin$   \\ 
 $4.7$  & $0.6$  &  $8$  & $0.68\arcmin \times 4.02\arcmin$   \\ 
 $2.25$  & $0.08$  &  $40$ & $1.42\arcmin \times 8.5\arcmin$  \\ 
  $1.25$  & $0.08$ &  $200$ & $2.6\arcmin \times 16.3\arcmin$ \\ 
\hline
\end{tabular}
\end{table}

\begin{table}
\caption{Measured flux densities for PSO J0309+2717 at different epochs. Notes on columns: (1) -- epoch in MJD; (2), (4) -- number of observations; (3), (5) -- measured flux densities at 4.7 and 8.2 GHz.}            
\label{table:fluxes}      
\centering                         
\begin{tabular}{c | c c | c c}       
\hline                
MJD  & $N_{4.7}$ & $\bar{S}_{4.7}$ & $N_{8.2}$ & $\bar{S}_{8.2}$ \\    
      &                       &      mJy            &                       &       mJy           \\
                          (1)  &  (2)  &  (3)  & (4)  &  (5)    \\

\hline  
\emph{epoch 1}  & \multirow{3}{*}{4}  & \multirow{3}{*}{$14 \pm 3$} & \multirow{3}{*}{4} & \multirow{3}{*}{--}\\  
58977 -  & & & & \\          
58981    & & & & \\
\hline 
\emph{epoch 2}  & \multirow{3}{*}{2}  & \multirow{3}{*}{$12 \pm 3$} & \multirow{3}{*}{2} & \multirow{3}{*}{--}\\ 
59022,  & & & & \\          
59026   & & & & \\
\hline 
\emph{epoch 3}  & \multirow{3}{*}{40}  & \multirow{3}{*}{$10 \pm 2$}  &  \multirow{3}{*}{45} & \multirow{3}{*}{$8 \pm 3$} \\ 
59041 - & & & & \\ 
59096 & &  & &  \\ 
\hline                                   
\end{tabular}
\end{table}

\section{Comparison with other high-redshift blazars}  

\begin{figure*}
\begin{minipage}{0.32\textwidth}
\centering
\includegraphics[width=\linewidth]{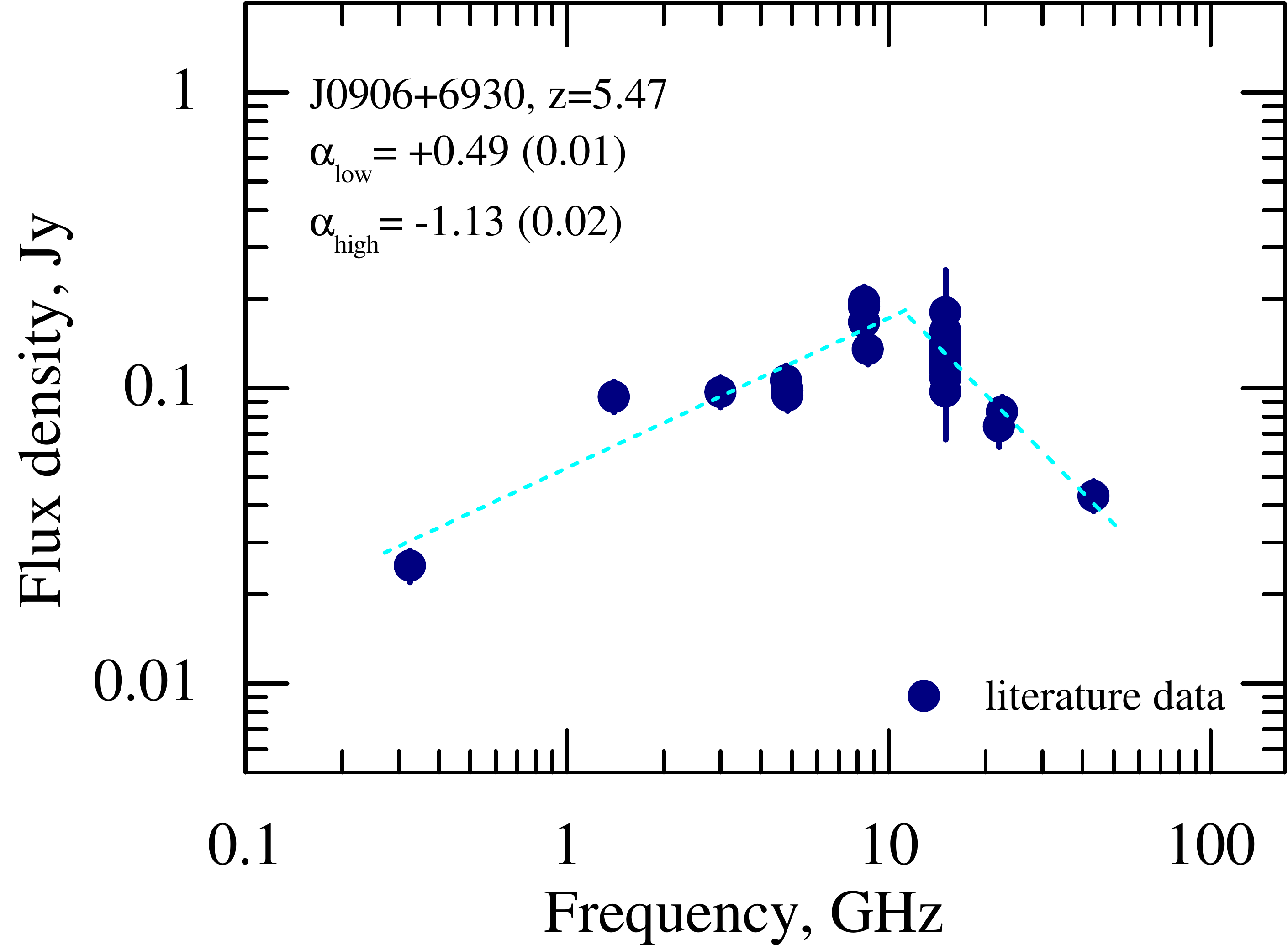}
\end{minipage}
\begin{minipage}{0.325\textwidth}
\centering
\includegraphics[width=\linewidth]{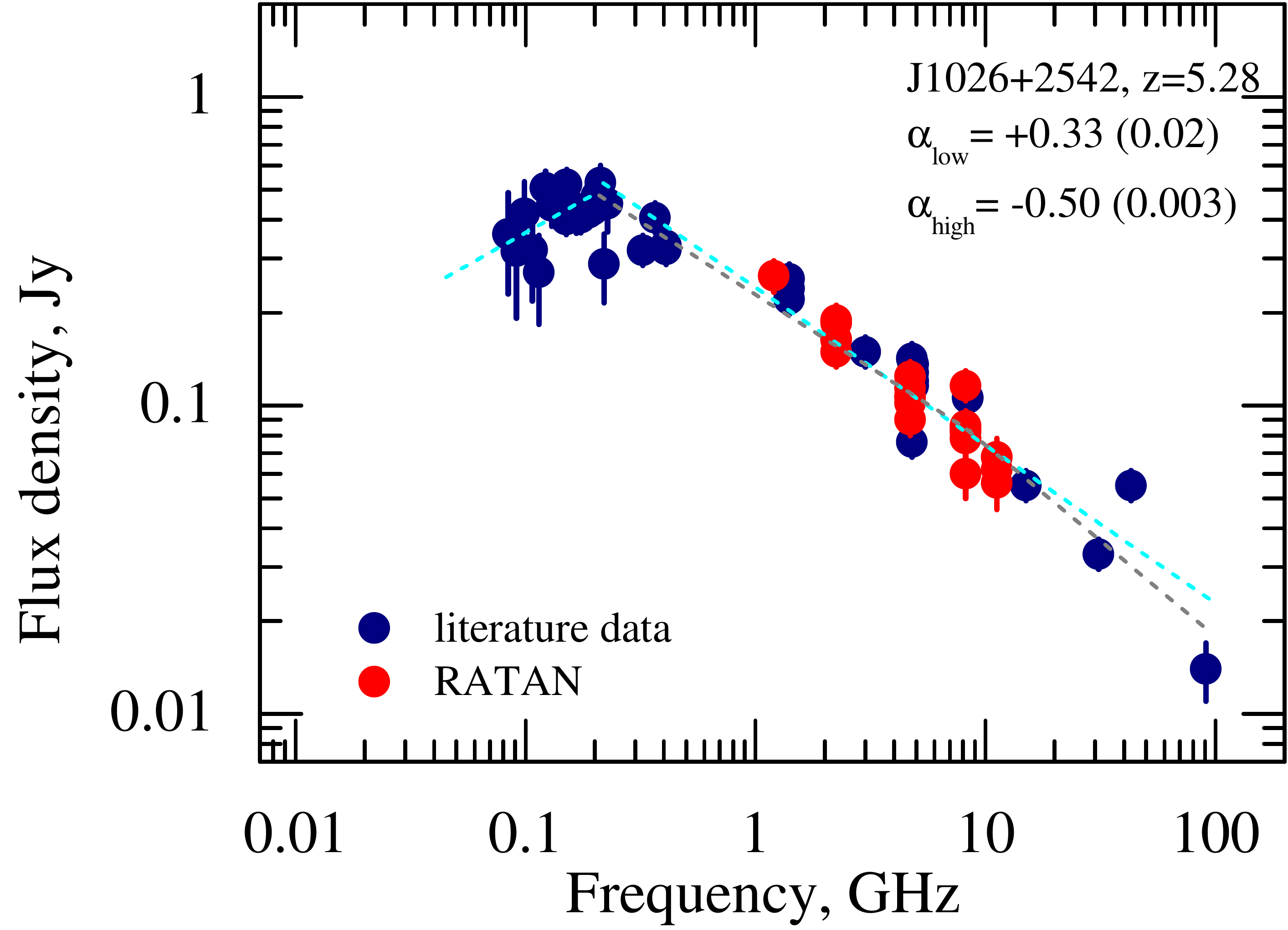}
\end{minipage} 
\begin{minipage}{0.33\textwidth}
\centering
\includegraphics[width=\linewidth]{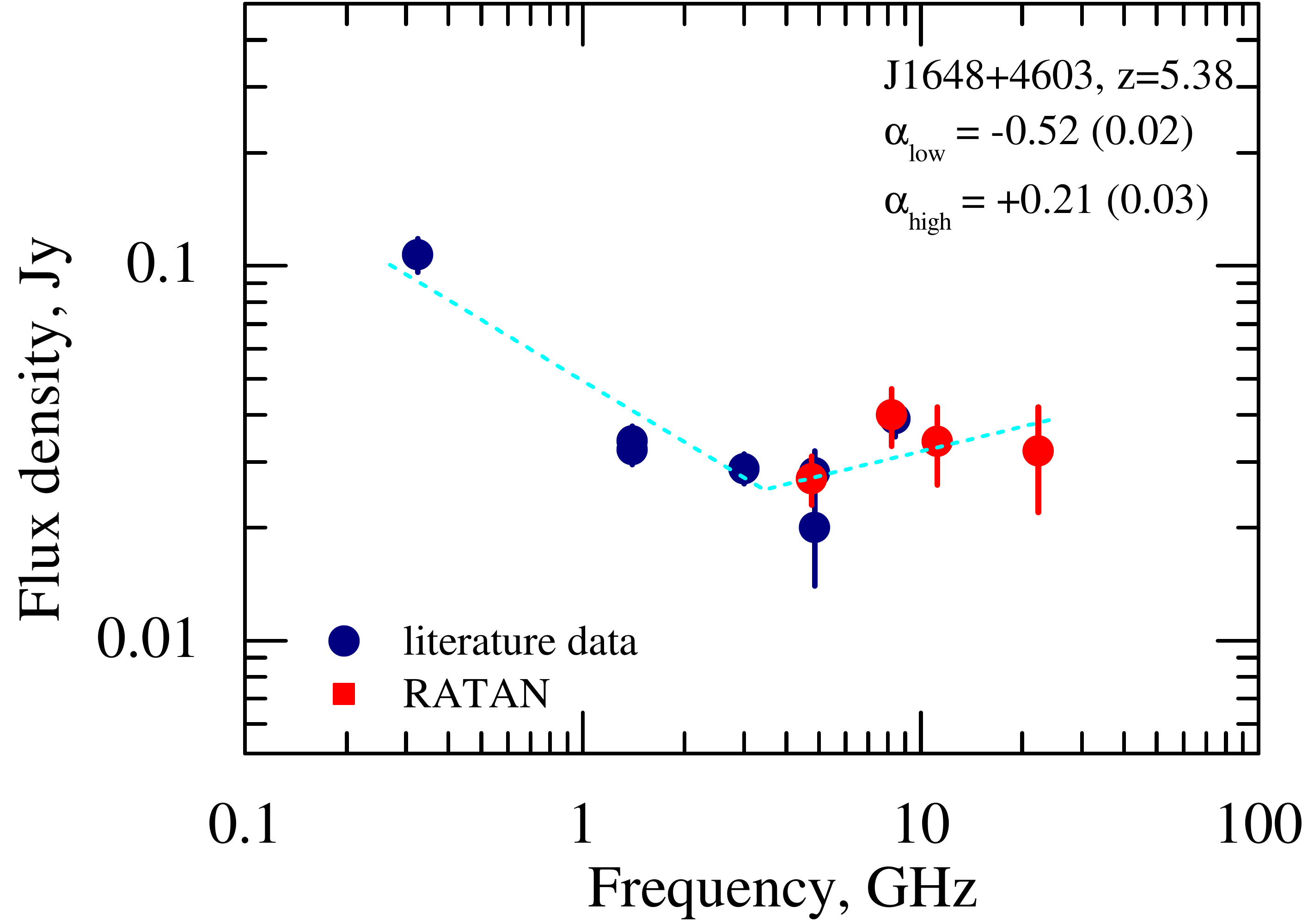}
\end{minipage}  
\label{Fig:figure3}  
\caption{Radio spectra of the three most distant blazars known at redshifts 5.47, 5.38, and 5.28. Blue points are from the literature, CATS, and VLASS; red points are RATAN-600 measurements.}
 \end{figure*}

We compared the radio spectrum of PSO J0309+2717 
with that of the three currently most distant blazars at $5.0 < z < 6.0$. 
They are BZQ J0906+6930 at $z$=5.47 \citep{2004ApJ...610L...9R}, 
BZQ J1026+2542 at $z$ = 5.28 \citep{2012MNRAS.426L..91S},
and NVSS J164854+460328 (J1648+4603 in what follows) at $z$=5.38 \citep{2019MNRAS.484..204C}.
The blazars have been classified in the literature based on the common criteria using
both radio and X-ray properties, and only for two of them,
BZQ J0906+6930 and BZQ J1026+2542, has Doppler boosting been confirmed with VLBI observations
that revealed a core--jet structure on parsec scales 
\citep{2015MNRAS.446.2921F,2017MNRAS.468...69Z, 2018A&A...618A..68F, 2020NatCo..11..143A}.
Their available radio spectra are presented in Fig~\ref{Fig:figure3},
where data from CATS\footnote{\url{www.sao.ru/cats}} \citep{1997ASPC..125..322V} and other literature data 
are plotted with blue circles, and RATAN-600 measurements for BZQ J1026+2542 
and J1648+4603 are presented as red circles.
Their radio properties are listed in Table~\ref{table:par}. 

The second most distant blazar, \textbf{BZQ J0906+6930}, is a bright radio source with a flux density at 1.4 GHz of 98 mJy at $z = 5.47$.
VLBI observations suggest a less powerful jet compared to blazars in general \citep{2020NatCo..11..143A}.
The peaked radio spectrum has a turnover frequency of $11.3\pm2.5$ GHz in the observer's frame or $73.1\pm16.1$ GHz
in the source rest frame, which makes BZQ J0906+6930 a high-frequency peaking object (HFP) candidate. 
We calculated the spectral indices as $\alpha_{low} = 0.49 \pm 0.01$ and $\alpha_{high} = -1.13 \pm 0.02 $.
Its light curve obtained at 15 GHz with the OVRO 40-m radio telescope 
demonstrates moderate variability $V = 0.3$ \citep{2017MNRAS.468...69Z}. 
At frequencies of about 4.8 and 8 GHz, the variability index is 0.06 and 0.18 over time-scales of 9 and 10 yr respectively (CATS data base).
A small number of radio measurements, covering about 20 yr of observations, and possible variability could mean that the radio spectrum of BZQ J0906+6930 is possibly not a peaked one.

The blazar \textbf{BZQ J1026+2542} is one of the brightest distant blazars in the radio domain with a flux density
 of 257 mJy at 1.4 GHz \citep{1998AJ....115.1693C}.
Its radio spectrum was obtained from simultaneous measurements
at five frequencies from 1.2 to 11.2 GHz with RATAN-600 in the period from
 2017 February--2019 August.
The flux-density measurements were obtained at 11.2, 8.2, 1.2 GHz with an S/N ratio
  $>5\sigma$ detection level and with S/N $>10\sigma$ at 4.7, 2.3 GHz (Table~\ref{tab:1026p25}).
During six observing epochs, the blazar was not detected at 22.3 GHz 
(the upper limit is about 11 mJy), and it was detected at 1.2 GHz only once in 2017 December.
The radio spectrum of BZQ J1026+2542 is peaked with the maximum $\nu_{obs}$=$0.21\pm0.02$ GHz
in the observer frame of reference ($1.32\pm0.13$ GHz in the source rest frame). The spectral indices below and above the peak frequency are $\alpha_{low} = 0.33 \pm 0.02$ and $\alpha_{high} = -0.50 \pm 0.03$ respectively. 
At frequencies above 8 GHz its spectrum gets steeper and can be described by two linear parts with $\alpha_{1}=-0.47 \pm 0.01$ and $\alpha_{2}=-0.63 \pm 0.01$ (Fig~\ref{Fig:figure3}).
The spectrum of BZQ J1026+2542 is well determined in the 0.084--91 GHz frequency range, and the measurements cover a period of more than 40 yr (according to the CATS data base).
The source demonstrates quite low variability over a time-scale of 13--14 yr: $V=0.11$ at about 8 GHz based on the VLA measurements in 2003--2006 \citep{2007ApJS..171...61H} and on the average RATAN-600 fluxes during 2017--2019; the variability index based on the RATAN measurements at 2.3--8.2 GHz is also estimated in the range of $V=0.11-0.21$. The radio spectrum below the peak is mainly represented by
quasi-simultaneous GMRT measurements, which are in good 
agreement with the other MHz measurements \citep{1996AJ....111.1945D,2017A&A...598A..78I}.
The peaked spectrum seems to be real for BZQ J1026+2542 since its variability is quite low and is similar to that of the known gigahertz-peaked-spectrum sources (GPS) \citep{2013AstBu..68..262M, 2019AstBu..74..348S, 2020arXiv200902750O}.
 
The radio spectrum of \textbf{J1648+4603} can be classified as complex based on the literature data   
\citep{1991ApJS...75.1011G,1996ApJS..103..427G,1997ApJ...475..479W,1997A&AS..124..259R,1998AJ....115.1693C,2003MNRAS.341....1M}
and RATAN-600 simultaneous measurements at 4.7, 8.2, and 11.2 GHz, with $\alpha_{low}$=$-0.52\pm0.02$ and $\alpha_{high}$=$0.21\pm0.03$.
Two linear fits give a value of 3.7 GHz for the frequency of the minimum flux density of the spectrum.
The flux densities measured with RATAN-600 at 4.7 and 8.2 GHz in 2020 are consistent within the uncertainties with the Green Bank 4.85 GHz measurements in 1986 November and 1987 October and with the VLA 8.4 GHz measurement in 1994--1995 \citep{1996ApJS..103..427G,2003MNRAS.341....1M}. 
This suggests the absence of significant flux variations of the source over long time-scales. 
The spectrum of J1648+4603 can be explained by a sum of two major components. One of them is flat or peaked and can be associated with a parsec-scale jet that dominates at frequencies greater than 5 GHz. The second one is steep at frequencies up to 1 GHz, representing synchrotron radiation of optically thin extended structures up to kiloparsec scales \citep{2002PASA...19...83K}. The spectral shape and variability properties for J1648+4603 are not determined robustly due to  the absence of systematic measurements.

As a result, we have found that two of the four currently known most distant blazars have peaked spectra, and one has a complex spectrum with a peaked part at high frequencies. PSO J0309+2717 with a formally flat radio spectrum has a hint to an ultra-steep spectrum at high frequencies if we consider the upper limit at 11.2 GHz.
 In this case, it is a possible peaked-spectrum source (MPS/CSS) with a turnover frequency $\nu_{obs} \leq $ 0.147 GHz.
 BZQ J1026+2542 has a GPS spectrum, based on a sufficient number of data points including simultaneous ones. For BZQ J0906+6930 the peaked type of spectrum is apparently caused by a compilation of scarce data points. 
Classical GPS sources have a steep or even ultra-steep spectrum in the optically thin emission mode. 
They are not variable, and their spectral shape and small angular size are explained by young ages.
However, the high-frequency research of the \citet{2011A&A...536A..14P} has shown that
a large fraction of sources having GPS spectra are associated with compact sources with beamed jets, usually with blazars. 

We compared the radio luminosities of the aforementioned blazars with those from a sample of $z\geq 3$ quasars 
(106 sources with $S >$ 100 mJy at 1.4 GHz in Sotnikova et al., in preparation).
The typical radio luminosity for that sample is $\sim 10^{27}$ -- $10^{28}$  W Hz$^{-1}$, 
which is consistent with the estimated values for the four most distant blazars.
Our calculated monochromatic radio luminosity for BZQ J0906+6930 ($P_{\nu} \sim3.9\times 10^{27}$ W Hz$^{-1}$) is almost the same as that of \citet{2018A&A...618A..68F}, where $P_{\nu} \sim4\times 10^{27}$ W Hz$^{-1}$ was obtained at frequencies of 2.3 and 8.6 GHz. 
On average, our derived values of radio luminosity at 4.7 GHz for the four blazars at $z > 5$ are typical and in good agreement with those for the sample of distant radio-loud quasars at $z > 4.5$ in \citet{2016MNRAS.463.3260C}.
The radio luminosity of low-redshift quasars is significantly lower; e.g. \citet{2019MNRAS.485.2710J} reported $P_{\nu} \sim$  $10^{23}$ -- $10^{24}$ W Hz$^{-1}$ for a sample of $z < 0.2$ obscured quasars.

\begin{table*}
\caption{Radio properties of the four most distant blazars currently known. Notes on columns: (1) source name; (2) redshift; (3),(4) peak frequency of the radio spectrum at observer's and rest frame; (5) type of the radio spectrum; (6),(7) spectral index below and above 4.7 GHz; (8) flux density at 4.7 GHz; (9) radio luminosity at 4.7 GHz.}  
\label{table:par}      
\centering                          
\begin{tabular}{c c c c c c c c c}        
\hline              
Source         & $z$ & $\nu_{obs}$ & $\nu_{int}$ & Type & $\alpha_{low}$ & $\alpha_{high}$ & $S_{4.7}$ & $P_{\nu}$ \\    
                   &        &  GHz              &             GHz &              &                        &                          & mJy                    & W/Hz  \\         
 \\ (1)  &  (2)  &  (3)  & (4)  &  (5)  &  (6)  &  (7)  &  (8)  &  (9) \\             
\hline                        
 PSO J0309+2717 & 6.10 & -- & --            & flat        & $-0.44\pm 0.10$ & $-0.51\pm 0.10$ & $12\pm3$ & $1.9\pm0.10 \times 10^{27}$  \\     
 BZQ J0906+6930 & 5.47 & $11.3\pm2.5$ & $73.1\pm16.1$  & peaked  & $0.49\pm 0.01$ & $-1.13\pm 0.02$  & $106\pm5$ & $3.9\pm0.01 \times 10^{27}$ \\      
 BZQ J1026+2542 & 5.28 & $0.21\pm0.02$ & $1.32\pm0.13$ & peaked  & $0.33\pm 0.02$ & $-0.50\pm 0.03$  & $108\pm10$ & $1.3\pm0.01 \times 10^{28}$  \\      
 NVSS J164854+460328 & 5.38 & -- & --  & complex & $-0.52\pm 0.02$ & $0.21\pm 0.03$ & $31\pm5$ & $1.5\pm0.01 \times 10^{27}$ \\       
\hline                                   
\end{tabular}
\end{table*}

\begin{table*}
\caption{RATAN-600 new measurements for BZQ J1026+2542 in the period of 2017--2019 and for J1648+4603 in 2020 August. Flux densities are given in units of mJy for different frequencies (22.3, 11.2, 8.2, 4.7, 2.25, and 1.25 GHz). *************** *** ***** ** ******.} 
\label{tab:1026p25}
\centering
\begin{tabular}{cccccccc}
\hline
 Source & MJD & $S_{22.3}$ & $S_{11.2}$ &  $S_{8.2}$ & $S_{4.7}$ & $S_{2.25}$ & $S_{1.25}$ \\
\hline
 \multirow{6}{*}{J1026+2542} & 57805 & $<6$ & $68 \pm10$ & $86 \pm10$ & $114 \pm10$ & $164\pm10$ & $<200$ \\
 & 57837 & $<6$ & $56 \pm10$ & $84 \pm10$ & $124 \pm10$  & $189\pm10$ & $<105$  \\
 & 58106 & $<12$ & $62 \pm10$ & $82 \pm10$ & $106 \pm10$ & $185\pm10$ & $263\pm30$ \\
 & 58226 & $<8$ & $68 \pm10$ & $78 \pm10$ & $107 \pm10$ & $149\pm10$ & $<190$ \\
 & 58320 & $<18$ & $<45$ & $116 \pm10$ & $102 \pm10$ & $<150$ & $<100$ \\
 & 58715 & $<17$ & $<21$ & $60 \pm10$ & $90\pm10$ & $163\pm10$ & $<100$  \\
 \hline
J1648+4603 & 59068 & $32 \pm10$ & $34 \pm8$ & $40 \pm7$ & $27 \pm4$ & $<30$  & $<35$  \\
\hline
\end{tabular}
\end{table*}

\section{Summary}

The newly discovered most distant blazar PSO J0309+2717 at a redshift $z = 6.1$
 was observed with the RATAN-600 radio telescope at three frequencies simultaneously: 4.7, 8.2, and 11.2 GHz. 
 It was detected at 4.7 GHz with S/N$ > 4$ (2020 May--September ) and at 8.2 GHz with S/N$ > 2.5$ (2020 July--September ). 
 The averaged flux-density over three epochs is $12.0\pm3$ mJy at 4.7 GHz and $8\pm3$ mJy at 8.2 GHz.
The flux density upper limit at 11.2 GHz is estimated as 3 mJy. 

Using the new RATAN-600 measurements along with literature data at 0.147, 1.4, and 3 GHz, we have revealed a flat radio spectrum with $\alpha_{0.147-8.2}$=$-0.51\pm0.1$. This result is in good agreement with the previous spectral index estimates from \cite{2020A&A...635L...7B} and \cite{2020A&A...643L..12S}. However, if we consider the upper limit at 11.2 GHz, the radio spectrum steepens at higher frequencies ($\alpha_{5.8-11.2} \leq -1.4 \pm 0.05$), in the same way as the spectral index of BZQ J1026+2542 steepens from $\alpha = -0.4$ to $\alpha = -0.7$ \citep{2013ApJ...777..147S}.

The light curve at 4.7 GHz obtained on a time-scale of four months exhibits moderate variability with $F_{var}=0.28\pm0.02$. Future long-term observations at higher frequencies will help us to assess the possible variability of the source. 

Comparison of the radio properties of PSO J0309+2717 with other distant blazars
 has revealed that their radio luminosities are similar at $\sim10^{27}$ W Hz$^{-1}$. 
This value is consistent with radio luminosities for high-redshift quasars at $z\geq 3$ (Sotnikova et al., in preparation) and $z > 4.5$ \citep{2016MNRAS.463.3260C}.

\section*{Acknowledgements}

    We thank the referee for providing useful suggestions and comments that significantly improved the paper.

   This work is supported in the framework of the national project ``Science'' by the Ministry of Science and Higher Education of the Russian Federation under the contract 075-15-2020-778.
   
   The observations were carried out with the RATAN-600 scientific facility.

      Observations with RATAN-600
      are supported by the Ministry of Science and Higher Education of the Russian Federation. 

     MM, TM, and VS acknowledge support through the Russian
     Government Programme of Competitive Growth of Kazan Federal University.
     
    This research has made use of the NASA/IPAC Extragalactic Database (NED),
    which is operated by the Jet Propulsion Laboratory, California Institute of Technology, 
    under contract with the National Aeronautics and Space Administration.


\section*{Data Availability}
The data underlying this article are available in the article.

\bibliographystyle{mnras} %
\bibliography{manuscript} %

\bsp	
\label{lastpage}
\end{document}